\documentclass[12pt,preprint]{aastex}
\usepackage{graphics}
\newcommand{\degree}{\hbox{$^\circ$}}

\newcommand{\ltsimeq}{\la}
\newcommand{\gtsimeq}{\ga}
\newcommand{\msun}{M$_{\odot}$}

\newcommand{\HII}{H~{\sc ii}}

\received{}
\revised{}
\accepted{}
\shortauthors{McQuinn et al.}
\shorttitle{The True Durations of Starbursts}

\begin{document}
\title{The True Durations of Starbursts: HST Observations of Three Nearby Dwarf Starburst Galaxies}
\author{Kristen B.~W.~McQuinn\altaffilmark{1}, 
Evan D.~Skillman\altaffilmark{1},
John M.~Cannon\altaffilmark{2},
Julianne J. Dalcanton\altaffilmark{3},
Andrew Dolphin\altaffilmark{4},
David Stark\altaffilmark{1},
Daniel Weisz\altaffilmark{1}
}

\altaffiltext{1}{Department of Astronomy, School of Physics and
Astronomy, 116 Church Street, S.E., University of Minnesota,
Minneapolis, MN 55455, \ {\it kmcquinn@astro.umn.edu}} 
\altaffiltext{2}{Department of Physics and Astronomy, 
Macalester College, 1600 Grand Avenue, Saint Paul, MN 55125}
\altaffiltext{3}{Department of Astronomy, Box 351580, University 
of Washington, Seattle, WA 98195}
\altaffiltext{4}{Raytheon Company, 1151 E. Hermans Road, Tucson, AZ 85706}
\begin{abstract}

The duration of a starburst 
is a fundamental parameter affecting the
evolution of galaxies yet, to date, observational
constraints on the durations of starbursts are not well established.
 Here we study the recent star formation histories (SFHs) of three nearby 
dwarf galaxies to rigorously quantify the duration of their starburst 
events using a uniform and consistent approach.
We find that the bursts range from $\sim 200 - \sim 400$ Myr in duration resolving 
the tension between the shorter timescales often 
derived observationally with the longer timescales derived from 
dynamical arguments. 
If these three starbursts are typical of starbursts in dwarf galaxies,
then the short timescales ($3 - 10$ Myr) associated with starbursts in previous
studies are best understood as ``flickering'' events which are simply
small components of the larger starburst. In this sample of three 
nearby dwarfs, the bursts are not localized events.
All three systems show bursting levels of star formation 
in regions of both high and low stellar density.
The enhanced star formation moves around the galaxy during the bursts
 and covers a large fraction of the area of the galaxy. 
These massive, long duration bursts can significantly affect the structure, 
dynamics, and chemical evolution of the host galaxy and can be the progenitors
of ``superwinds'' that drive much of the recently chemically enriched material
from the galaxy into the intergalactic medium. 

\end{abstract} 


\section{A Perspective on Starburst Durations \label{intro} }

\subsection{Duration: A Fundamental Property of a Starburst \label{property} }
Starbursts are finite episodes of intense star formation which have been observed in dwarf, spiral, and interacting galaxies both in the local universe and in galaxies at high redshift (z~$\gtsimeq$~1) where they are thought to be more common \citep[e.g.,][]{Thompson05}. The concentrated star formation of a burst may have a significant impact on the surrounding environments through both energy and mass transfer altering not only the structure and composition of the host galaxy but also the local intergalactic medium (IGM) outside of the host galaxy \citep[e.g.,][]{Strickland00, Martin02}. For low mass dwarf systems\footnote{Using the $\textit{K}$-band luminosity as a tracer of stellar mass, a dwarf galaxy is defined by M$_{K} > -21$ \citep{GilDePaz03}}, the effect of a starburst on the galaxy and its evolution can be dramatic given the large energy output of starburst (i.e., supernovae and mass loss driven winds) compared with the dwarfs' smaller potential wells \citep{Dekel86}. 

The impact of starbursts on galactic structure may have been more important in the early star-forming universe. Recent studies have estimated that 6\% of locally detected galaxies show evidence of starbursts based on their H$\alpha$ emission line \citep{Lee06}. The percentage of galaxies exhibiting starburst characteristics increases to 15 \% for galaxies detected at \textit{z} $= 1$ \citep{OConnell05}. Although there is a significant selection effect against detecting distant non-bursting galaxies, a greater fraction of star formation occurring in bursts in the early universe is expected due to the increased frequency of galaxy interactions \citep[a known starburst trigger;][]{Kennicutt87} and the greater gas fractions found at these earlier epochs. While bursting, the energy output of a starburst dominates the total luminosity of the host galaxy. As a group, local starbursts (\textit{z} $\ll$ 1) provide $\sim$10\% of the radiant energy production and $\sim20-25$\% of all high mass star formation in the local universe \citep[e.g.,][]{Heckman98, Brinchmann04, Lee06}. 

There are a number of competing theories on the role starbursts play in shaping the host galaxy and the impact of starbursts on the host galaxy's evolutionary track. Many authors \citep[e.g.,][]{Loose86,Silk87,Davies88,Mayer01a, Mayer01b, Mayer06} have postulated that starbursts may be a phase of evolution morphing dwarf irregulars (dIrrs) into dwarf spheriodals (dSphs) or ellipticals (dEs). The role of starbursts in this evolutionary scenario may be to assist in the depletion of gas while accounting for the loss of metals to the IGM and to perhaps play a somewhat lesser role in shaping the global structure and dynamics of the galaxy. Using a larger set of dwarf galaxy observations, \citet{vanZee04} resolve inconsistencies between the angular momentum distribution of dEs and of dIrrs \citep{vanZee01} which could not be previously accounted for if dIrrs morph into dEs. Along a different line of thought, \citet{Salzer02} suggested starbursts make up a distinct class of dwarfs and that the progenitors of starburst dwarf galaxies are not typical dwarf galaxies. Starbursts would therefore play a more active role in determining the structure of the host galaxy. Despite the uncertainties around the galaxy classification scheme of starbursts \citep{Kennicutt98} and the details of its impact \citep{Heckman05}, it is commonly accepted that starburst events are a significant phenomenon affecting the host galaxy and its environs.

The strength of a starburst can be characterized by three fundamental parameters: its size relative to the host galaxy, the relative intensity of the star formation, and the duration of the burst. If bursts are short in duration, the intensity and size may determine how much impact the burst has on the galaxy, but, if the durations are long, then the overall affect on a galaxy's evolution and structure may be dramatic whether the burst is relatively large, strong or otherwise. The strength of a starburst affects many other processes within the host galaxy. For example, starbursts can drive galactic winds, a potentially important enrichment and heating process of the IGM and a possible natural explanation for the galaxy mass-metallicity relationship \citep[e.g.,][and references therein]{LeeH06}. Understanding the duration parameter is fundamental in understanding the relationship between starbursts and galactic winds \citep[e.g.,][]{Recchi06}. It is plausible that longer durations lead more naturally to 'superwinds' \citep{Cooper08} therefore successful modeling of galactic winds and the evolution of dwarf galaxies requires an accurate measurement of the duration of a starburst \citep[e.g.,][]{Spaans97, Romano06}. Similarly, models of outflows from bursting systems (i.e., blow-outs and blow-aways) which show different characteristics for an instantaneous burst and for bursts where the energy input is temporally extended \citep[e.g.,][]{MacLow99, Ferrara00} need an accurate determination of burst duration. In addition, the burst duration directly affects other inferred parameters such as the chemical yield and the age of the galaxy \citep{Kobulnicky97}. Likewise, interpreting galaxy luminosity functions is dependent upon the burst duration since it is the duration that directly affects the slope of the faint end of the luminosity function as burst galaxies spend more time at low luminosities than at high luminosities \citep[e.g.,][]{Hogg97}. 

\subsection{Long vs. Short Durations \label{longvshort} }
Starbursts are thought to be a short-lived phenomenon from an astronomical perspective. Two early papers in the study of dwarf starburst galaxies \citep{Searle72, Searle73} inferred from the composition and colors of these ``isolated extragalactic \HII$~$regions" that the galaxies are ``undergoing intermittent and unusually intense bursts of star formation" \citep{Searle73}. These authors reason from observations that the blue colors of starburst galaxies are a result of flashes of star formation over timescales of $10^{8}$ yr superimposed over a lower average rate of star formation from the birth of the galaxy $10^{10}$  yr ago. The shorter timescale indicates that these periods of intense star formation are discontinuous and must be finite or cyclical in nature. The duration of a single burst could be dictated primarily by fuel limitations or by energy and mass transfer feedback mechanisms from the stellar populations of the burst \citep[i.e., supernovae and stellar winds,][]{Thornley00, Ferguson98} or by both.

Although starbursts are well-accepted as a short-lived phenomenon, measurements of dwarf galaxy starburst durations are scarce and contradictory, even at the level of distinguishing between possibly artificial delineations of short (``self-quenching"; $\simeq$ 5 Myr) and long ($\simeq 100$ Myr) bursts. Observational studies of the emission from Wolf-Rayet (W-R) stars \citep{Schaerer99, Mas-Hesse99} find burst durations of $\sim 2 - 4$ Myr. In the study of W-R galaxies, the strength of the He II $\lambda$4686 line relative to the H$\beta$ line is correlated to the total number of W-R stars. A high ratio is interpreted as a short burst producing many W-R stars simultaneously, while a long burst would distribute the W-R stars in time resulting in a relatively weak $\lambda$4686 line (i.e., that longer duration bursts tend to dampen spectral features and lead to lower burst parameter strengths). Although the W-R star measurements typically sample a very small area of the galaxy, these studies assume that the spectroscopic observations sample the entire burst population.

On a larger scale, \citet{Thornley00} studied 27 starburst galaxies using infrared line emission ratios to confirm the presence of massive stars (M $\simeq 50-100$ \msun) in starburst sites. These authors find a declining radiation field hardness inferred from neon ionic abundance ratios, implying that the most massive stars producing the Ne lines have already evolved off the main sequence. The same study used the declining ratio of the bolometric to Lyman continuum luminosity ratio as a timescale for lower mass stars to evolve off the main sequence thereby constraining the IMF and invoking a short duration of $1 - 10$ Myr to explain the emission ratios. They further suggest stellar winds and supernovae disrupt the gas distribution and destroy the starburst environment limiting durations to $<10^{7}$ yr rather than a longer burst governed by a gas consumption timescale. 

The studies of star clusters in nearby bursting galaxies have used a number of different methods to measure the duration of starbursts. Tremonti et al. (2001) found ages of $1 - 8$ Myr for the central clusters in NGC 5253 by fitting starburst models to ultraviolet (UV) spectra, which are sensitive to the young, high-mass, stellar population. They report that the field was populated by stellar clusters dissolving on timescales of 10 Myr. Cluster dissolution would create a bias against observing older clusters constraining the measurement of a burst duration to star formation within an individual stellar cluster. Harris et al.~(2004) also report a duration of 10 Myr in star cluster regions in NGC 5253 and in NGC 3077, based on theoretical population synthesis models. In contrast, durations of at least 100 Myr in NGC 5253 have been found by combining the ages of the diffuse stellar population with the bright stellar clusters determined using optical imaging, and H$\alpha$ and H$\beta$ spectra \citep{Calzetti97}. Separately, \citet{Ostlin03} used UV and optical observations of ESO 338-IG04 coupled with spectral evolutionary synthesis models to place a lower bound on burst durations in star clusters at $\sim 40 - 50$ Myr.

A similar tension between young ($\sim$10 Myr) ages in localized star clusters and older ($\sim100$ Myr) ages in the broader star forming regions has also been seen in NGC 1569. Young star clusters and areas of enhanced star formation in NGC 1569 have been dated with ages $\ltsimeq 30$ Myr \citep{Hunter00} using integrated UVI colors, although analysis of the optical color magnitude diagrams of the field stars in NGC 1569 yielded elevated star formation over the last 100 Myr \citep{Greggio98}. Starburst scenarios that incorporate star formation inside and outside stellar clusters find durations based on the UBI colors of the clusters and diffuse light of $10 - 100$ Myr corresponding to up to 10 times the theoretical crossing time \citep{Meurer00}. Synthesis modeling in the H$\alpha$ equivalent width and B$-$V and U$-$B color plane by \citet{Lee06} converges on bursts with durations of $50-100$ Myr.

On the theoretical side, \citet{Tosi89} modeled dwarf irregulars in the Local Group by comparing observational color-magnitude diagrams with theoretical simulations and report a burst duration of 5 Myr for the WLM galaxy. Ferguson et al.~(1998) also posit that feedback from the burst (stellar winds and supernovae (SNe)) quench future star formation so that only short-duration bursts are possible in very faint (B $\gtsimeq$ 24) dwarf galaxies. These authors use the physical conditions, degree of `burstiness', the detectability of low luminosity galaxies and the fiducial 10 Myr duration to model faint dwarf galaxies, supported by the reasoning that at longer times the type II SNe produced by the burst would heat the interstellar medium preventing any further star formation. In another simulation, \citet{Stinson07} describe bursting star formation as driving heated gas into a galaxy's halo which quenches star formation. The same gas cools and is later accreted back onto the galaxy triggering another burst in star formation. These authors report an oscillatory period of star formation of $\sim 300 - 400$ Myr for lower mass to higher mass halos.

In this paper, we present uniform measurements of the durations of starbursts in three nearby bursting dwarf galaxies: NGC 4163, NGC 4068, and IC 4662. The durations are explicitly determined using detailed recent ($\ltsimeq 1$ Gyr) star formation histories derived from resolved stellar populations seen with the Hubble Space Telescope (HST). We apply the same photometric techniques and treatment of differential extinction to the three galaxies. Our emphasis on uniformity permits a direct comparison of the star formation histories from each galaxy while minimizing systematic uncertainties. \S\ref{data} describes the galaxy sample, the observations and the data reduction process, \S\ref{burst} discusses the analysis to determine the burst durations, \S\ref{implications} compares our results with past studies and discuss the implications of longer lasting bursts. The last section (\S\ref{conclusions}) gives a brief summary of our results. A follow-up publication will extend the analysis to a larger sample of nearby dwarf starburst galaxies.

\section{The Data and the Analysis Techniques\label{data}}
\subsection{The Sample of Galaxies and Their Observations\label{galaxies}}

We have selected three dwarf galaxies, NGC 4163, NGC 4068, and IC 4662, for study. These galaxies were identified by \citet{Karachentsev06} as starburst\footnote{Definitions of a starburst system vary significantly. Some definitions use simple calculations about the gas consumption timescale signifying a finite episode of star formation. Others use a comparison between current bursting star formation rates to the star formation rate associated with normal or quiescent star formation activity in the past \citep[cf.,][]{Scalo86, Kennicutt98}. There are starburst definitions that look at the strength of ultraviolet and H$\alpha$ emission to identify current massive star formation \citep{Lee06}. Bursts have been inferred from the star formation per unit area in specific regions of a galaxy or in star clusters \citep[e.g.,][and references therein]{Heckman05}. While no definition can be applied indiscriminantly to galaxies at varying redshifts, each definition is useful in a specific context in identifying systems undergoing these intense and unsustainable periods of star formation.} galaxies
based on the properties of their color magnitude diagrams (CMDs) and the population distribution of their blue and red helium burning (BHeB and RHeB) stars. Identifying a burst using its CMD is based on a global qualitative view of stellar populations of the galaxy rather than on a quantitative analysis. The BHeB stars are intermediate mass stars with helium burning cores \citep{Dohm-Palmer02}. The lifetime of this evolutionary stage is relatively short; the typical age of a BHeB star in this region of the CMD ranges from $\sim5 - 600$ Myr. At ages great than $\sim600$ Myr, the BHeB branch merges into the red clump. A galaxy with a constant star formation rate (SFR) will have a distribution of BHeBs that gradually increases from bright to faint magnitudes with a tendency towards redder colors at the lower magnitudes. In contrast, the CMDs presented by \citet{Karachentsev06} show an over-density or clustering of BHeBs at intermediate magnitudes closer to the bluer color of the main sequence stars (MS) indicative of the increased recent star formation found in a burst. The population of bright main sequence stars is another indicator of recent star formation. 

\begin{figure}
\plotone{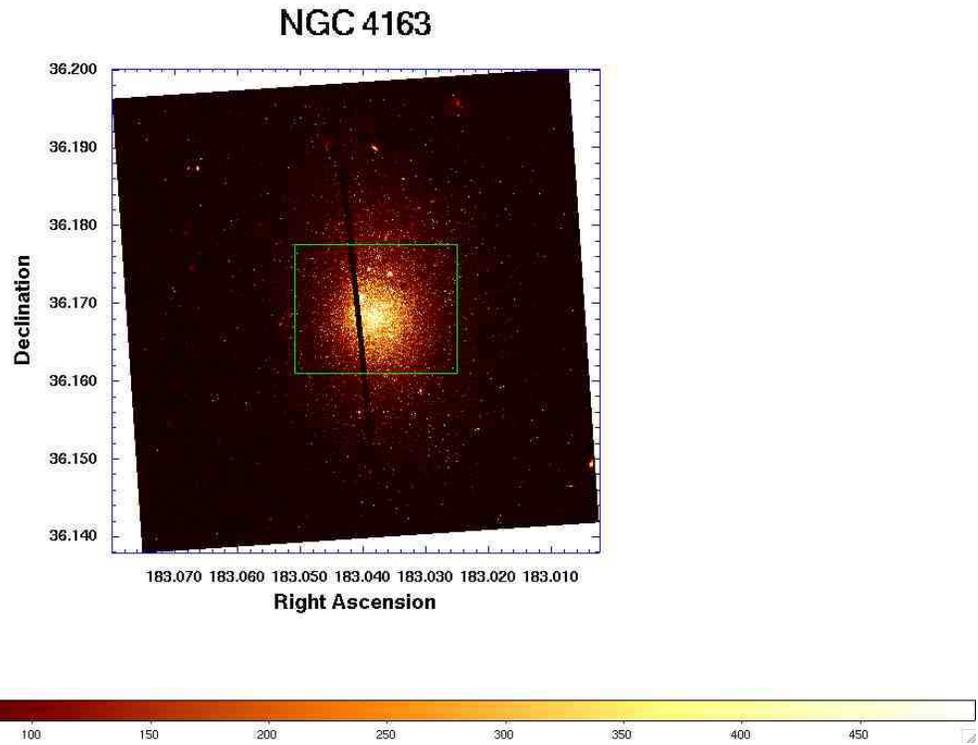}
\caption{ACS F606W image (V filter) of NGC 4163 using 1200 s of exposure. The green box encloses the area of higher stellar density where additional artificial stars were applied. SFH analysis was performed on the entire image and compared with the SFHs derived separately for the region of higher stellar density inside the box and the region of lower stellar density outside the box.}
\label{fig:ngc4163}
\end{figure}

\begin{figure}
\plotone{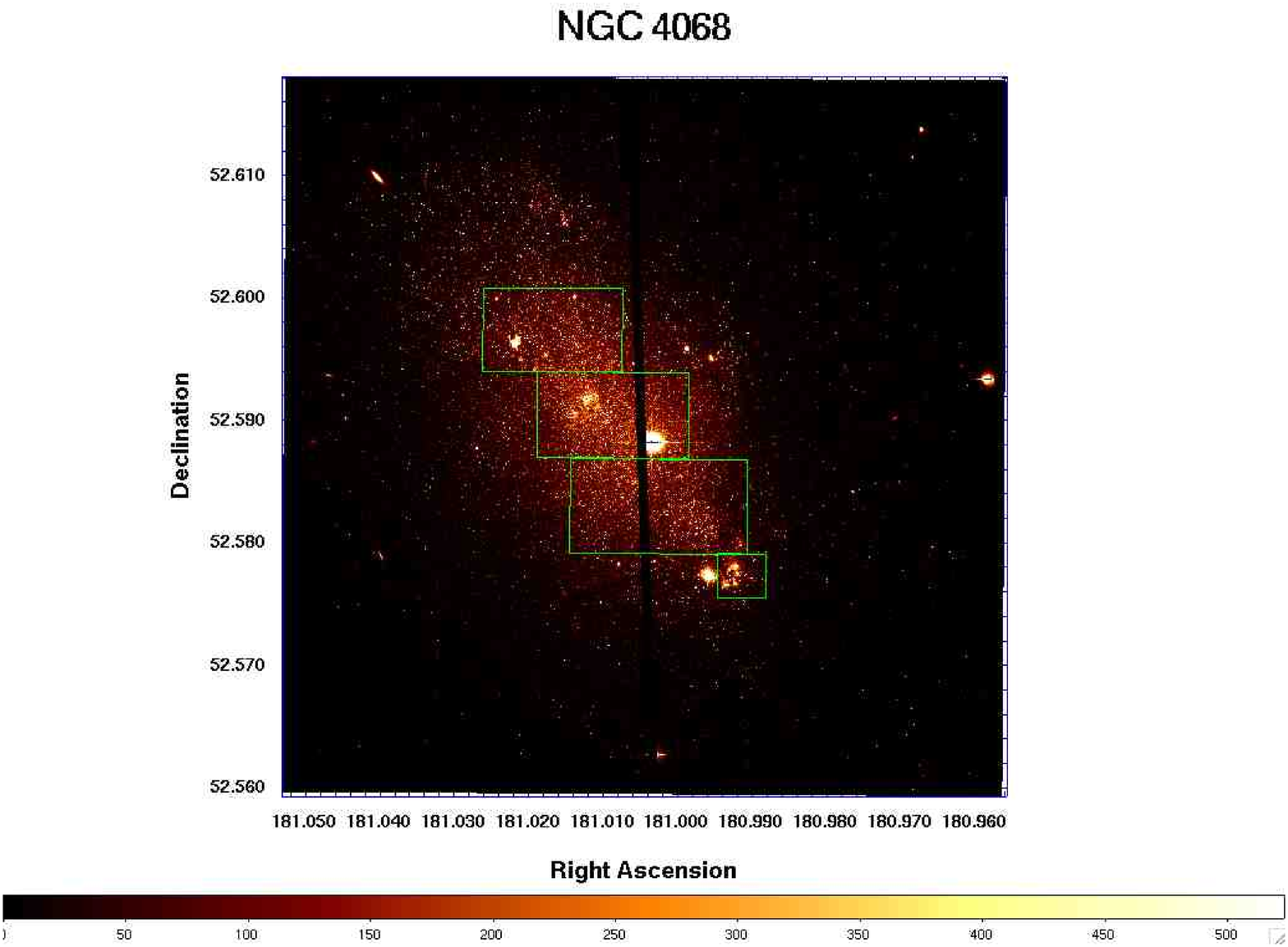}
\caption{ACS F606W image (V filter) of NGC 4068 using 1200 s of exposure. The green boxes encloses the area of higher stellar density where additional artificial stars were applied. SFH analysis was performed on the entire image and compared with the SFHs derived separately for the region of higher stellar density inside the boxes and the region of lower stellar density outside the boxes.}
\label{fig:ngc40668}
\end{figure}

\begin{figure}
\plotone{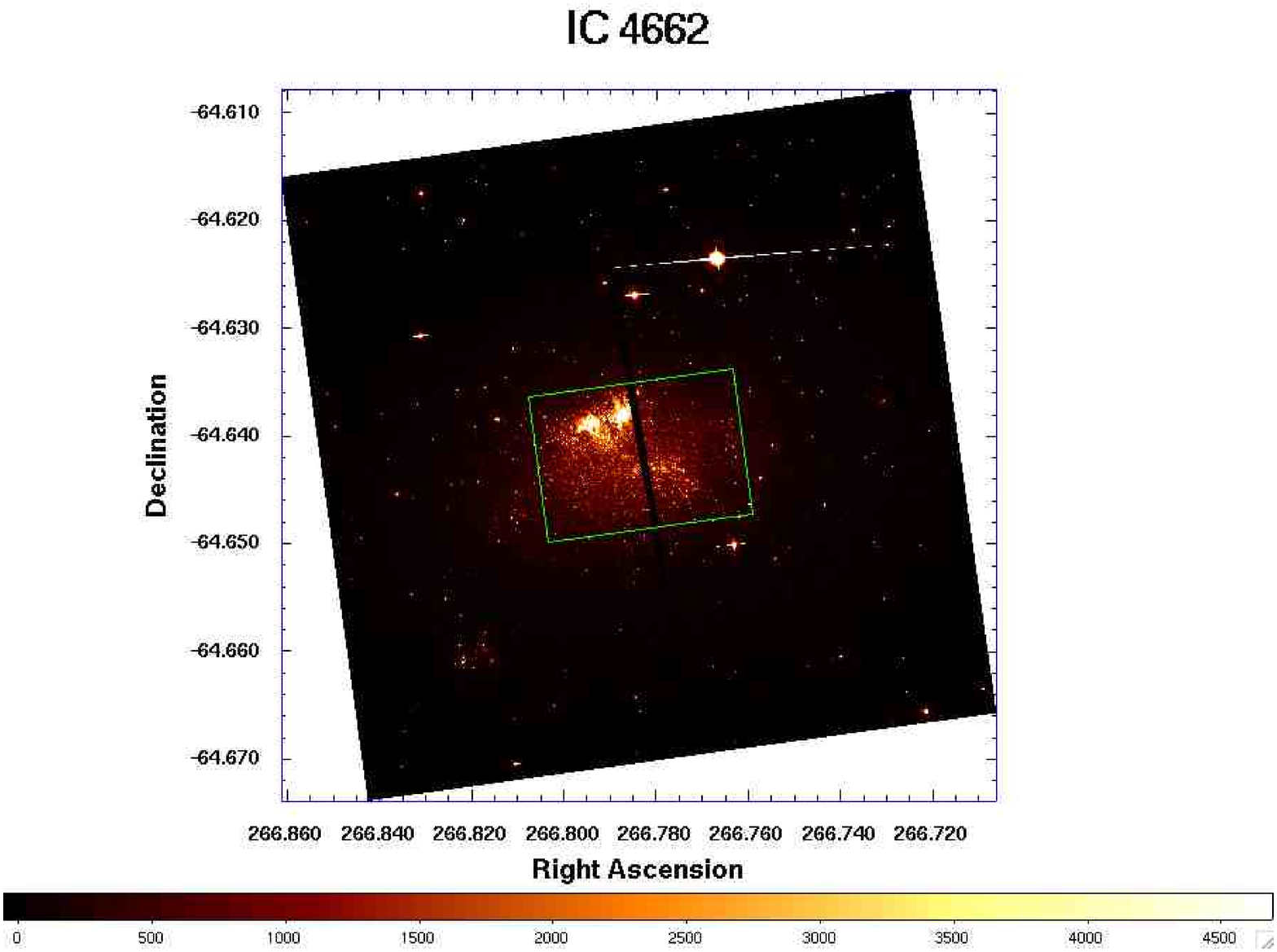}
\caption{ACS F606W image (V filter) of IC 4662 using 1200 s of exposure. The green boxes encloses the area of higher stellar density where additional artificial stars were applied. SFH analysis was performed on the entire image and compared with the SFHs derived separately for the region of higher stellar density inside the boxes and the region of lower stellar density outside the boxes.}
\label{fig:ic4662}
\end{figure}

The observations of all three galaxies were originally observed as part of program HST-GO-9771 \citep[PI:~Karachentsev,][]{Karachentsev06} and were retrieved from the Space Telescope Science Institute archive. The observations consist of 1200 s F606W and 900 s F814W images of each galaxy obtained using the Advanced Camera for Surveys (ACS) Wide Field Channel (WFC) on the Hubble Space Telescope. The images were cosmic-ray split and were cosmic-ray cleaned and processed by the standard ACS pipeline.  The observation details as well as other basic parameters, such as the distance and brightness of the galaxies, are summarized in Table~\ref{tab:galaxies}. The F606W images are shown in Figures~\ref{fig:ngc4163}$-$\ref{fig:ic4662} for NGC 4163, NGC 4068, and IC 4662 respectively. Areas of high surface brightness typically associated with higher stellar density and active star formation can be identified in each image.

\subsection{Photometry and Artificial Stars \label{photometry}}

Photometry was performed on the pipeline processed, cosmic ray cleaned 
images (CRJ files) using the ACS module of the DOLPHOT photometry package \citep{Dolphin00}. The photometry output of DOLPHOT includes several parameters, defined in \citet{Dolphin00}, which characterize the type of point source measured, the amount of crowding in its surrounding environment, the quality of the measurement, etc. Using these parameters, the photometry output file was filtered to select point sources identified as well-recovered stars with a minimum signal-to-noise ratio of 5 and $|$F606$_{sharp}+$F814$_{sharp}| \leq 0.39$ where a sharpness value of 0 characterizes a perfectly fit stellar PSF while a sharpness value of $\pm0.3$ in one wavelength represents a well-fit star.

The snapshot observations of the three galaxies reach a photometric depth comparable to the red clump with an absolute V magnitude of $\sim0 - 1$ given the distance moduli to the galaxies. This photometric depth is sufficient to rigorously determine the most recent (t $\ltsimeq 1$ Gyr) star formation history but lacks the necessary completeness at faint magnitudes to provide more than a rough characterization of the lifetime SFH. A photometric depth of at least M$_{V}~=~2$ is needed to accurately reconstruct the complete SFH of a galaxy \citep{Dolphin02}. As an assessment of accuracy of our method, a previous study of the Local Group galaxy Leo A with photometry of similar depth \citep{Tolstoy98} found an overall SFH that was later verified by \citet{Cole07} with deep photometry covering the oldest MS turnoffs. The ancient star formation history can be constrained using the RGB stars found above the observations' limits of these data, but the SFRs have additional uncertainties because of the covariance between time bins \citep{Dolphin02}. These uncertainties affect the rate within a particular time bin at large look-back times, but the total amount of star formation over the 14 Gyr and the averages are robustly derived numbers determined from our observational data \citep[e.g.,][]{Dohm-Palmer97}. 

\begin{figure}
\plotone{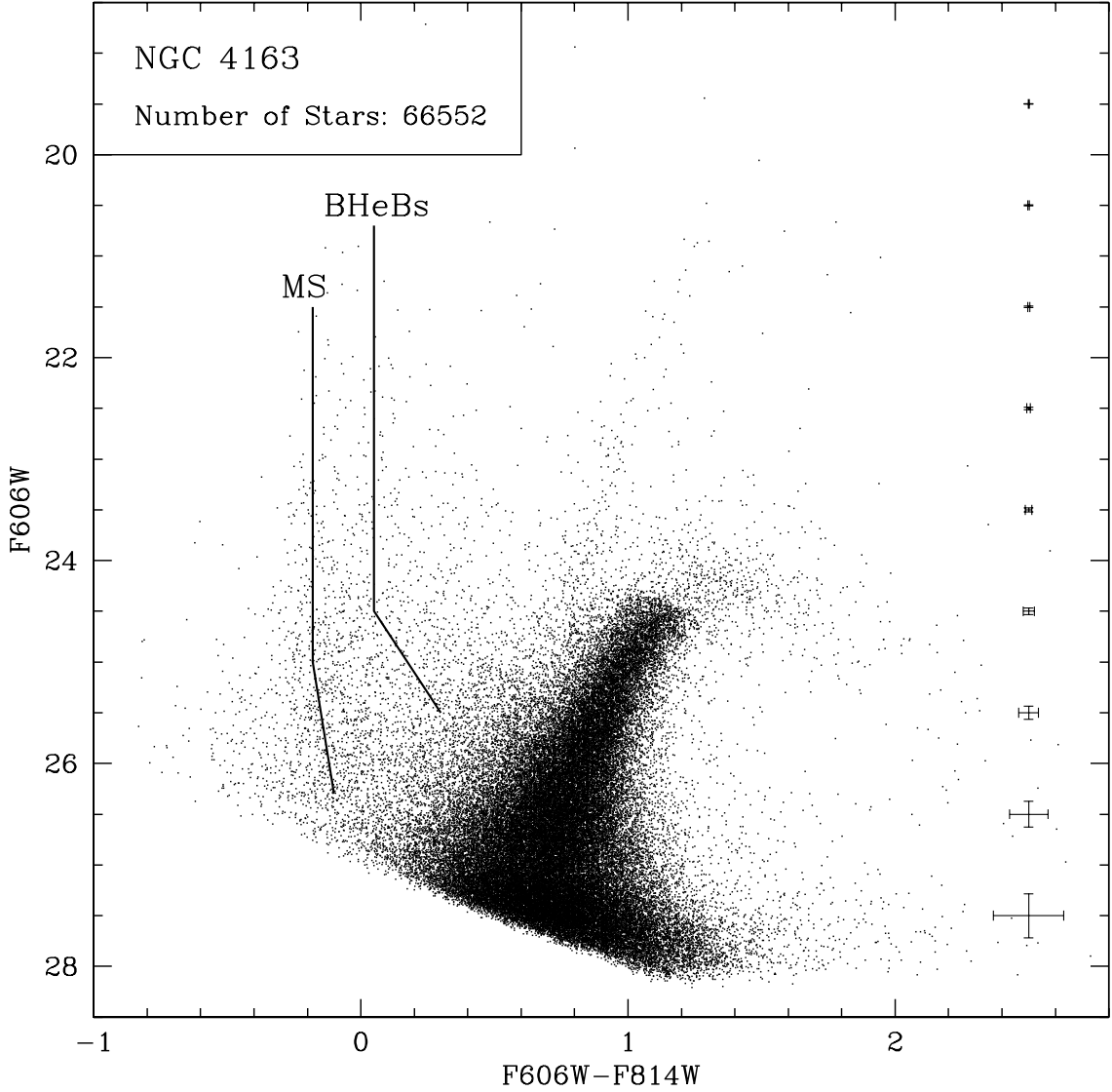}
\caption{The full field V vs. V$-$I color magnitude diagram for NGC 4163 contains over 66,000 stars. The TRGB, main sequence branch, and BHeB stars are identifiable in the CMD as well as the AGB stars and red supergiants. The distinct BHeB branch at V$-$I$\sim0$ spanning a V magnitude range of $~21$ and fainter distinguishes this galaxy as a starburst system. There is $\ltsimeq$ 0.1 mag of foreground extinction not corrected for in the CMD. The data are plotted to a signal-to-noise level of 5 which is approximately the 50\% completeness limit.}
\label{fig:cmd_ngc4163}
\end{figure}

\begin{figure}
\plotone{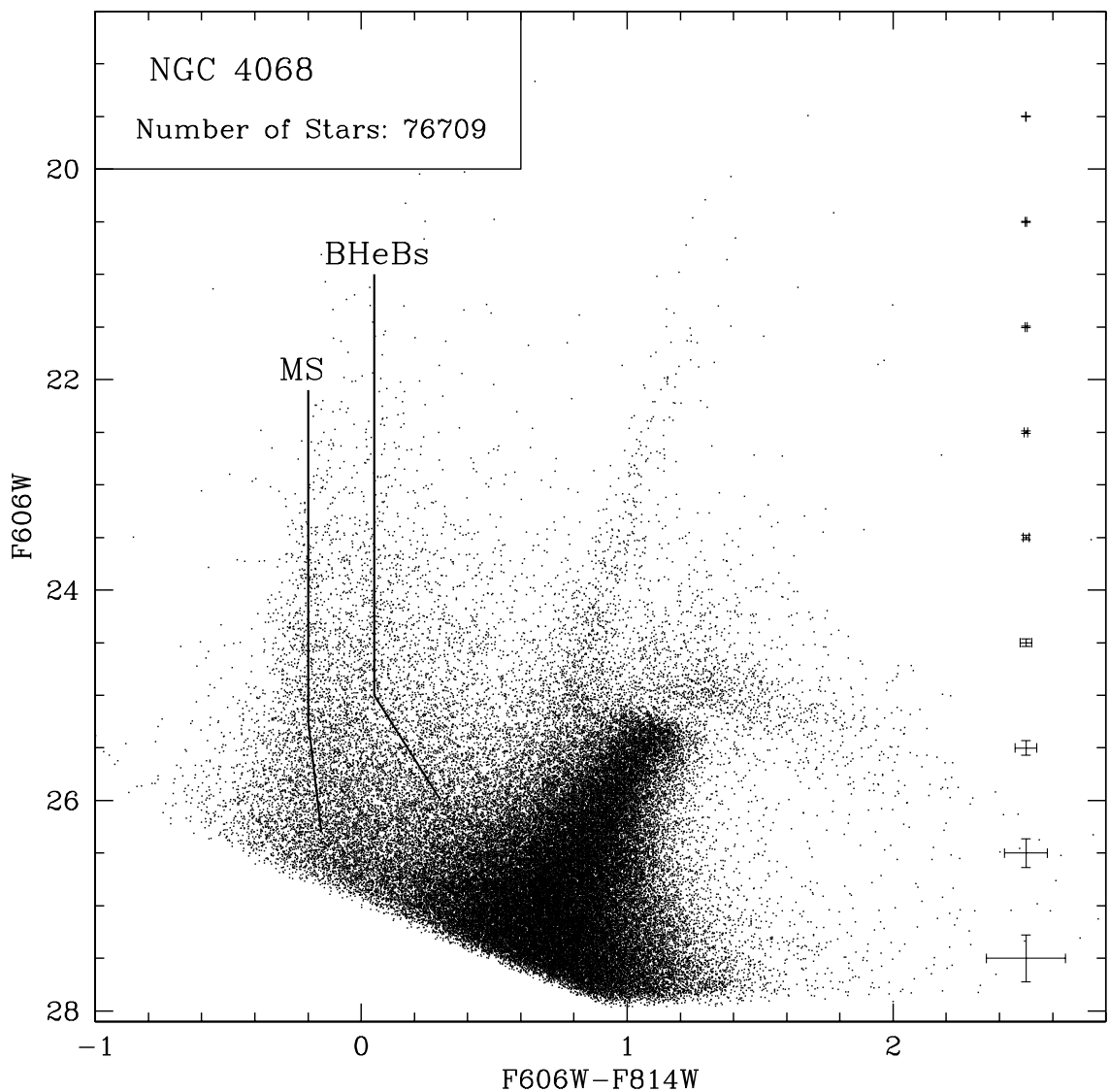}
\caption{The full field V vs. V$-$I color magnitude diagram for NGC 4068 contains over 76,000 stars. The distinct branch of BHeB stars at V$-$I$\sim0$ spanning a V magnitude range of $~21$ and fainter is more defined than NGC 4163 suggesting a larger burst in this galaxy. There is $\ltsimeq$ 0.1 mag of foreground extinction not corrected for in the CMD. The data are plotted to a signal-to-noise level of 5 which is approximately the 50\% completeness limit.}
\label{fig:cmd_ngc4068}
\end{figure}

\begin{figure}
\plotone{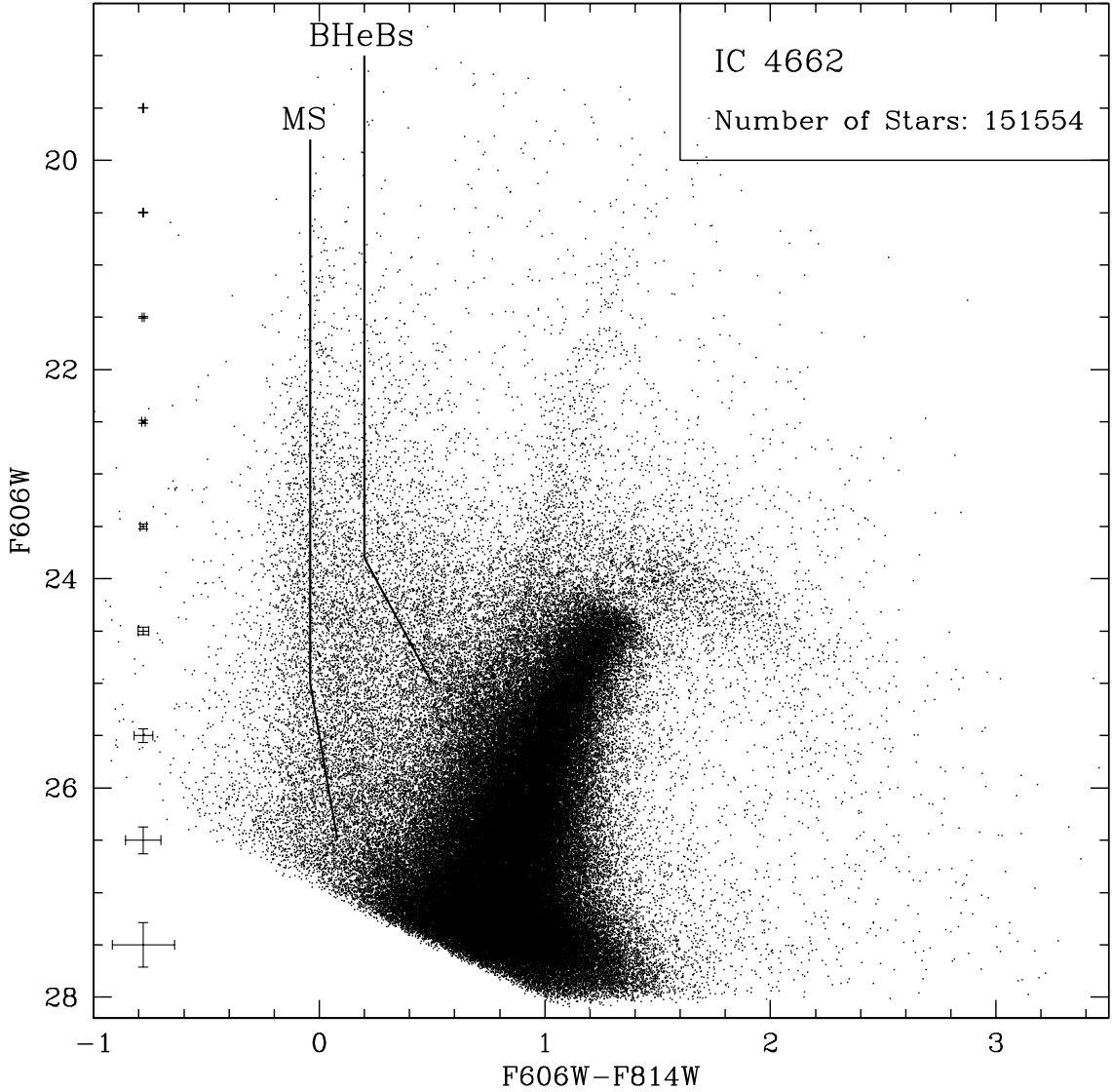}
\caption{The full field V vs. V$-$I color magnitude diagram for IC 4662 contains over 150,000 stars; the largest number of stars recovered photometrically in our sample. The TRGB, main sequence branch, and BHeB stars are clearly identifiable in the CMD and well populated. The distinct BHeB branch is slightly redward of the other two galaxies at V$-$I$\sim0.25$ spanning a V magnitude range of $~19$ and fainter, also distinguishing this galaxy as a starburst system. The main sequence and BHeB stars are blended at V magnitudes fainter than $~24$ due to the higher amount of foreground extinction ($\ltsimeq$ 0.5 mag) found in the field of view which is not corrected for in the CMD. The data are plotted to a signal-to-noise level of 5 which is approximately the 50\% completeness limit.}
\label{fig:cmd_ic4662}
\end{figure}

The CMDs resulting form our photometry for the three galaxies are shown in Figures~\ref{fig:cmd_ngc4163}$-$\ref{fig:cmd_ic4662}. Typical photometric errors per magnitude bin are shown in each of the CMDs in each figure, and the loci of the MS and BHeB populations are indicated by the superimposed lines. All three CMDs show MS and BHeB branches which are broadened to some extent. This broadening is due to three factors: photometric errors, crowding, and differential extinction. In the areas of highest stellar density, crowding will affect photometric accuracy \citep[e.g.,][]{Greggio98,Dohm-Palmer02}. The crowding causes an additional photometric uncertainty in the MS and BHeB branches (e.g., $\ltsimeq$ 0.03 mag at V$\sim24.5$) which are underestimated in the average photometric errors presented in the CMDs. We chose to include the point sources in areas of crowding with higher photometric noise rather than eliminate a fraction of the stars in which we are most interested. Both Galactic foreground differential extinction, when present, and differential extinction within the galaxy (see \S\ref{SFHs}) will broaden the evolutionary sequences. Two of the three - NGC 4163 and NGC 4068 - have little foreground Galactic extinction \citep[A$_{R}$ = 0.05 mag and 0.06 mag, respectively,][]{Schlegel98} lying well off the Galactic plane. The third galaxy, IC 4662, lies closer to the Galactic plane with a Galactic latitude \textit{b} $= -17.8$\degree. We find an A$_{V}$ value of $\sim0.3$ mag, somewhat higher than the Galactic extinction found by \citet{Schlegel98} of A$_{R}$ = 0.19 mag and $\simeq 0.2-0.4$ mag of differential extinction internal to IC 4662. The additional blending of the evolutionary sequences due to differential extinction in this galaxy is explicitly accounted for in CMD fitting program (see \S\ref{SFHs}). While these three factors create MS and BHeB branches that are somewhat blurred in the CMDs, one can still distinguish the separation between the two branches and the broadening does not affect the reconstruction of the star formation history.

The CMD of NGC 4163 (Fig.~\ref{fig:cmd_ngc4163}) contains the fewest number of recovered stars (66,000) and shows the least populated BHeB branch of the three galaxies, but has clearly undergone recent star formation. The more populated BHeB branch of NGC 4068 (Fig.~\ref{fig:cmd_ngc4068}) indicates a stronger burst. The BHeB branch in IC 4662 (Fig.~\ref{fig:cmd_ic4662}) is the most blended with the main sequence due to the higher foreground extinction and differential extinction present in the field of view of this galaxy. There is a small amount of foreground contamination seen in the CMD at bright magnitudes at a $V-I$ color of $\simeq 0.8$. It is the closest of the three galaxies with the deepest observations and contains the largest number of stars (150,000) in our sample.

\begin{figure}
\plotone{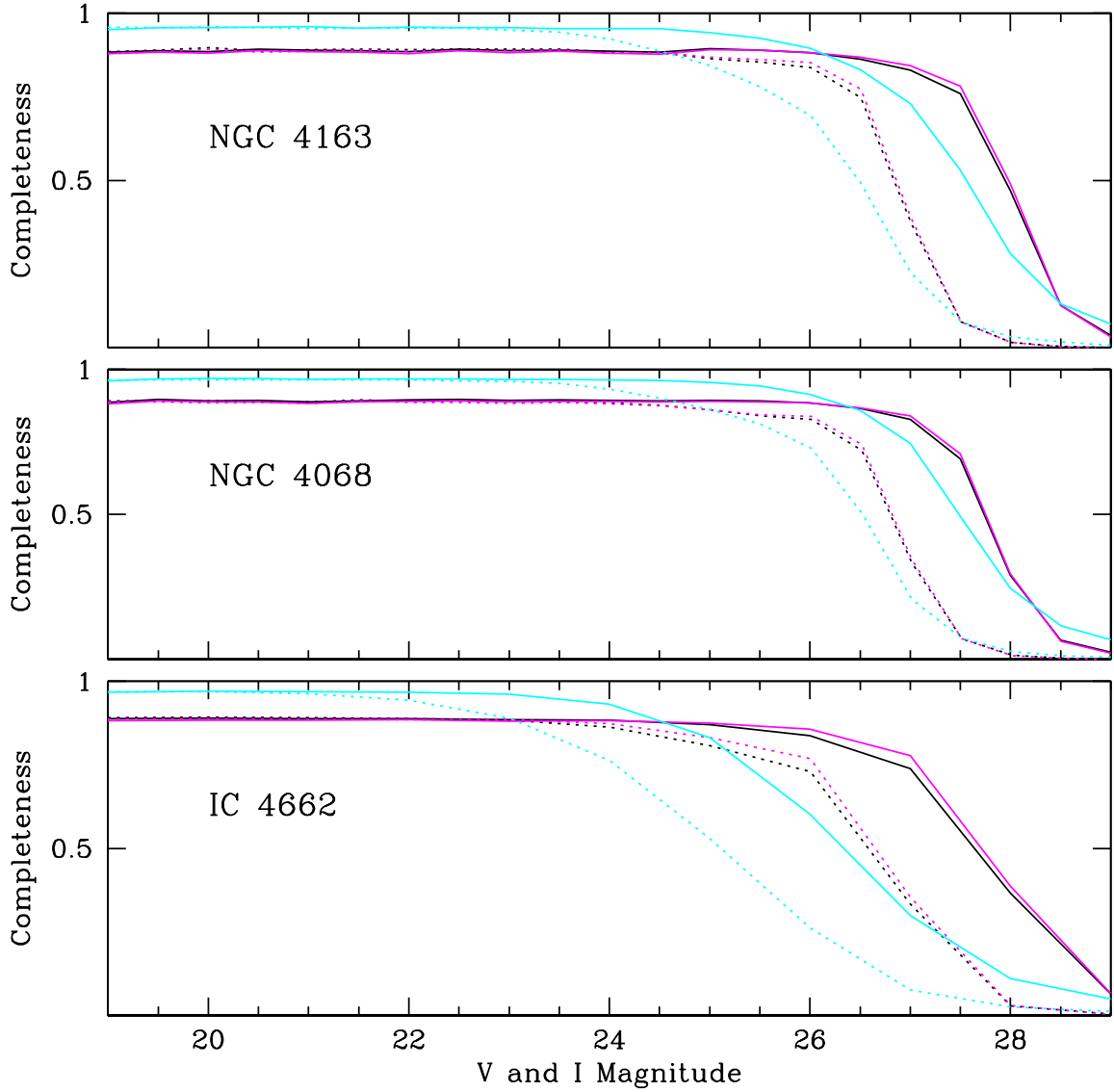}
\caption{Completeness as a function of magnitude for the three galaxies. The V (solid lines) and I (dashed lines) magnitude completeness limits are shown for the global solution (black), the lower density region (magenta), and the higher density region (cyan). The region of higher density reaches the confusion limit at bright magnitudes which artificially elevates the completeness limit.} 
\label{fig:completeness}
\end{figure}

To quantify the completeness of the data and the photometric uncertainty, we conducted artificial star tests both on the global image of each galaxy and on selected regions of each galaxy (regions highlighted in Figures~\ref{fig:ngc4163}-\ref{fig:ic4662}). Taken together, the two regions in each galaxy cover the entire field of view and were selected based on the relatively uniform surface brightness (i.e., stellar density) within each region. Applying false star tests to individual regions was useful in a number of ways. First, it allowed for a detailed SFH to be reconstructed on regions of more uniform characteristics which could then be compared to the global SFH results for a galaxy. Secondly, it allowed for a more precise determination of the location of a burst within a galaxy and facilitated a comparison of the SFRs in different regions. Thirdly, it served as one of the many tests we performed on the photometry probing the fidelity and accuracy of our results. We present completeness plots for the regions of lower and higher stellar density as well as for the global solution for the three galaxies in Figure~\ref{fig:completeness}. The completeness plateaus at $\simeq88\%$ in the global and lower density regions due to bad pixels and cosmic rays as the observations were obtained during single HST orbits (CRSPLIT = 2). The completeness for the region of higher stellar density is not easily quantified as the confusion limit is reached in these areas artificially increasing the completeness at bright magnitudes; as an upper limit, the completeness limit must be below the $88\%$ found for the less crowded regions. The CMDs for each galaxy (Figs.~\ref{fig:cmd_ngc4163}$-$\ref{fig:cmd_ic4662}) are plotted to a signal-to-noise ratio of 5 which corresponds approximately to the 50\% completeness limits. The final artificial star lists were filtered using the same parameters applied to the photometric output.

\section{The Burst Measurements \label{burst}}

\subsection {Star Formation Histories \label{SFHs}}

\begin{figure}
\plotone{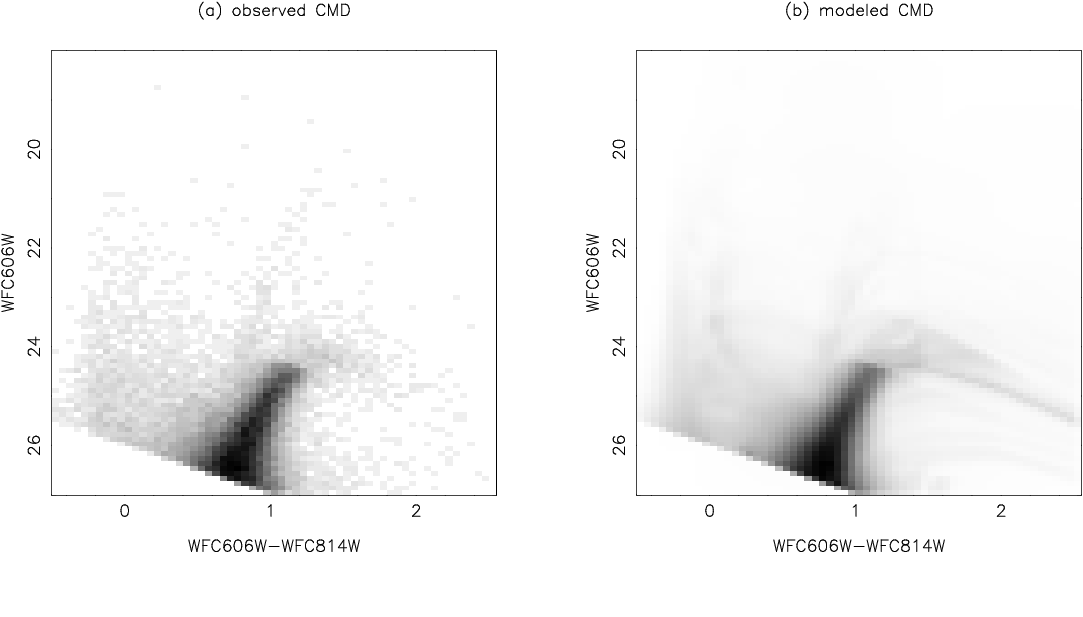}
\caption{The Hess diagram of the NGC 4163 observations and the best-fit synthetic Hess diagram of the galaxy.} 
\label{fig:synth_ngc4163}
\end{figure}

\begin{figure}
\plotone{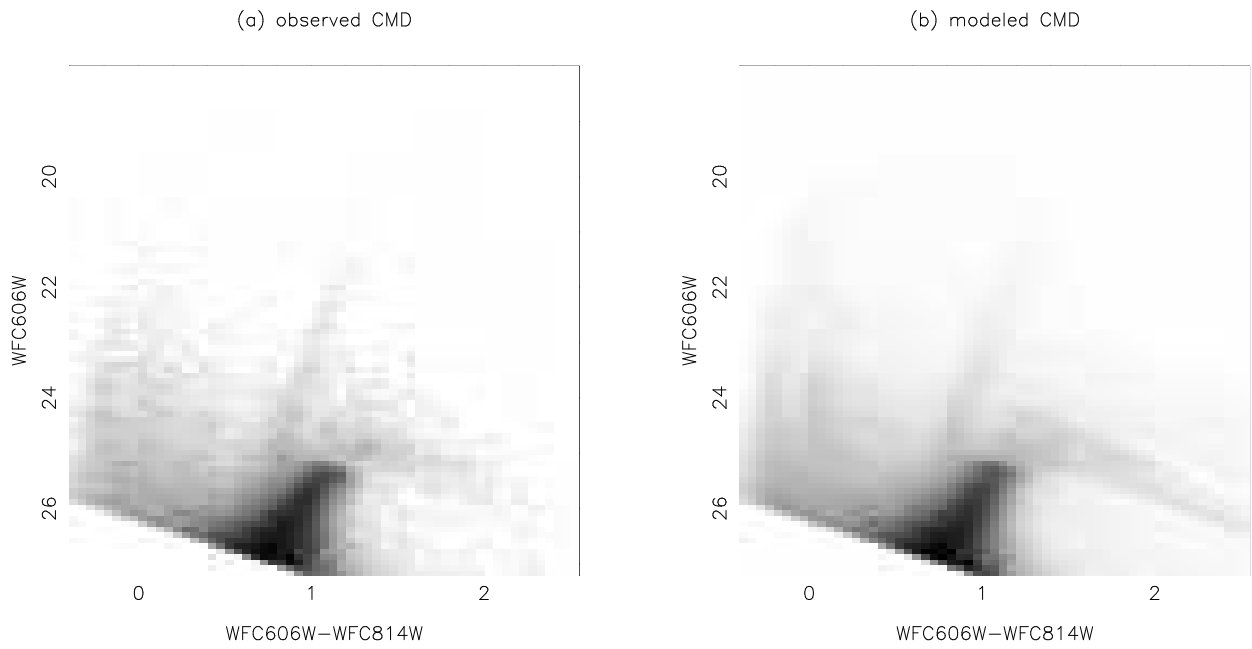}
\caption{The Hess diagram of the NGC 4068 observations and the best-fit synthetic Hess diagram of the galaxy.}
\label{fig:synth_ngc4048}
\end{figure}

\begin{figure}
\plotone{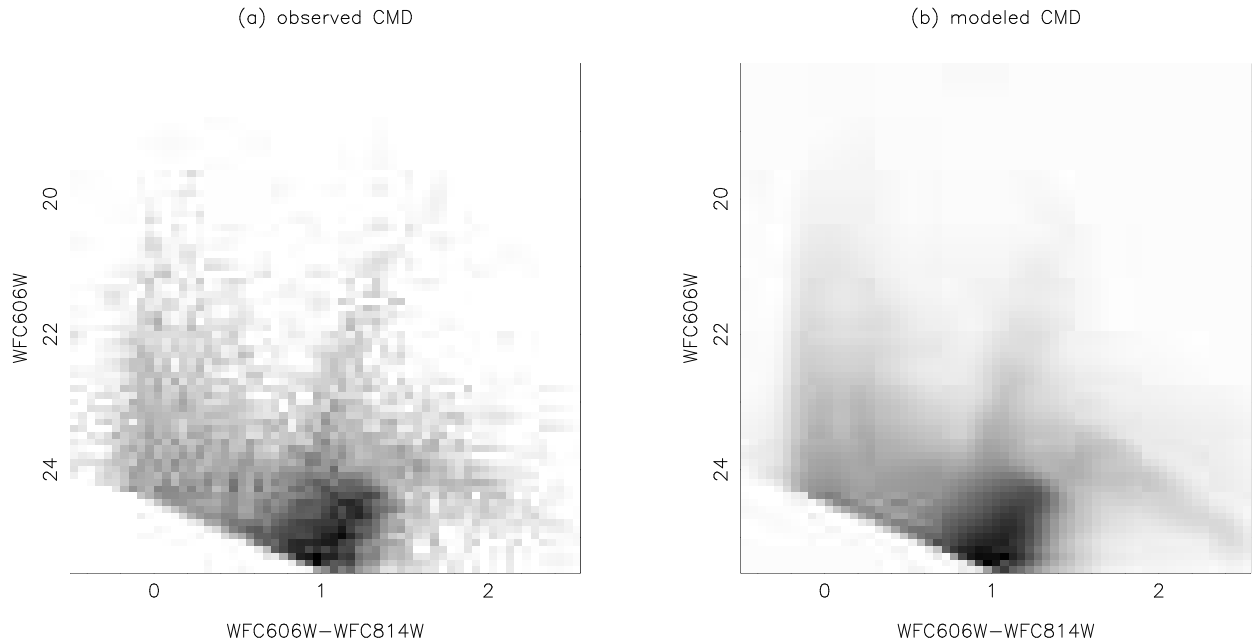}
\caption{The Hess diagram of the IC 4662 observations in the region of highest surface brightness and the best-fit synthetic Hess diagram of this region of the galaxy. Note the spread in the Red Giant Branch evolutionary track due to higher photometric errors from crowding in this part of the galaxy. There is also a significant amount of new star formation as evidenced by the well-populated Main Sequence and Blue Helium Burning branches. One of the signatures of differential extinction can also be seen in the partial blending of these two branches.}
\label{fig:synth_ic4662_hsb}
\end{figure}

\begin{figure}
\plotone{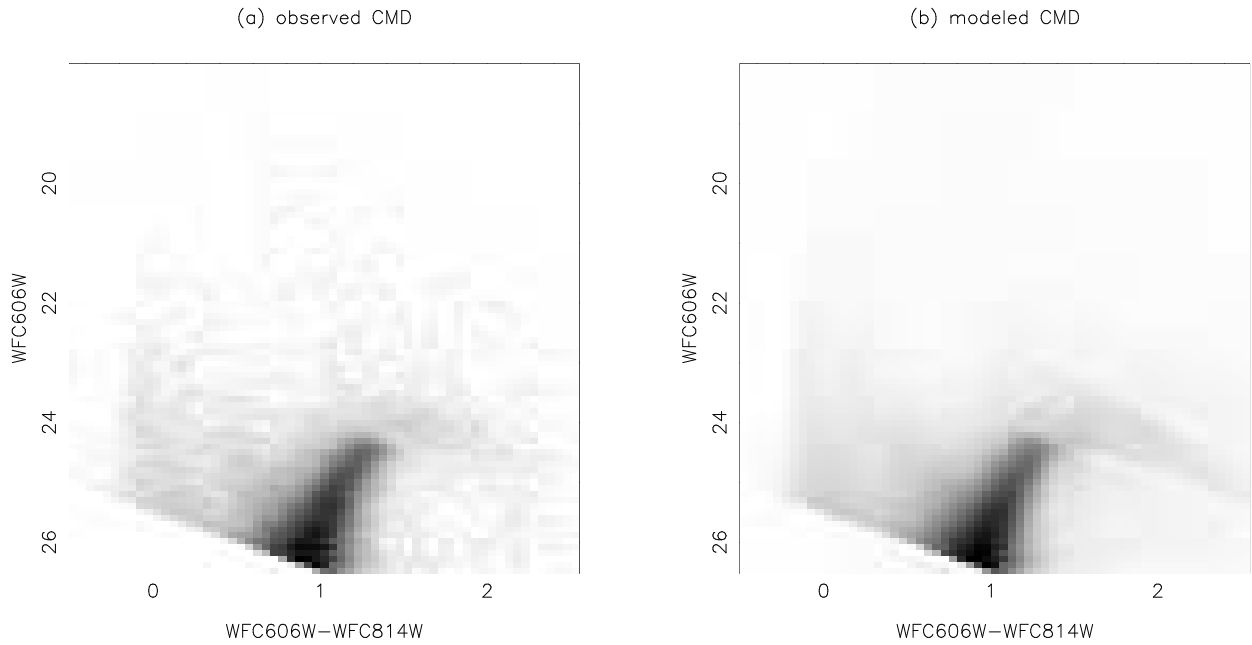}
\caption{The Hess diagram of the IC 4662 observations in the region of lowest surface brightness and the best-fit synthetic Hess diagram of this region of the galaxy. Note the tighter Red Giant Branch evolutionary track than in Figure~\ref{fig:synth_ic4662_hsb} due to lower photometric errors in this less crowded region of IC 4662.}
\label{fig:synth_ic4662_lsb}
\end{figure}

The star formation histories of the three galaxies were reconstructed using the color-magnitude diagrams. The photometry, observational errors, and incompleteness (i.e., artificial star recovery fractions) were used along with the stellar evolutionary models of \citet{Marigo07} in the SFH numerical method MATCH \citep{Dolphin02}. This CMD fitting program constructs a synthetic CMD based on the observed CMD and varies the metallicities and ages of the stellar populations. How well the observed and modeled CMDs match is quantified with a effective $\chi^{2}$ parameter \citep{Dolphin02} and reflects the likelihood of the SFH derived from the model CMD to be the true SFH of the observed galaxy. The $\chi^{2}$ per degree of freedom for our analysis was 1.3 in all cases except in the high surface brightness regions of IC 4662 where we achieved our best fit of 1.1, indicative of excellent fits. The modeled CMDs are presented as Hess diagrams in Figures~\ref{fig:synth_ngc4163}$-$\ref{fig:synth_ic4662_lsb} alongside the observed CMDs at the same axis scale for comparison. The Figures show that the models characterize all of the features of the CMDs quite well.  The best-fit synthetic CMD of the galaxy is based on a SFR as a function of time and metallicity which is the most likely SFH of the galaxy given our inputs and models \citep[i.e.,][]{Dolphin03,Weisz08}.

Numerous tests of the numerical method \citep[e.g.,][]{Dolphin03} have shown that the results robustly determine the SFR as a function of time and metallicity SFR(t,Z). The best modeled CMD fitting the observed data uses the most probable combination of distance to the galaxy and foreground extinction which can be directly compared with literature values. Given the large amount of gas present in a starburst galaxy needed to fuel the star formation, most starburst galaxies exhibit signs of differential extinction in their CMDs which blends the branches of stellar evolution. Therefore, differential extinction is applied within the CMD fitting program to the young stars in each galaxy decreasing linearly from A$_{v}$=0.5 for stars younger than 40 Myr to A$_{v}$=0.0 for stars older than 100 Myr \citep{Dolphin03}.

We applied several tests on the data to measure the effects extinction and photometric broadening may have on the SFH solutions in all parts of the CMDs. For example, selecting point sources with a more tightly constrained sharpness parameter (i.e., $|$F606$_{sharp}+$F814$_{sharp}| \leq 0.27$) or  point sources outside areas of highest crowding, did not change the SFH profiles or the duration of the burst derived. Likewise, assuming a higher level of extinction in the field of view than is actually thought to be present did not change the SFH profile or the burst duration. While these tests may suppress the absolute values of the SFR, we found that, regardless of the crowding of point sources in regions of higher stellar density and differences in foreground extinction present in our sample, the recent SFHs and the starburst duration measurements were robustly derived and our results remained the same. 

The errors from the fitting program quantify the systematic uncertainties in the SFHs due to variations of the real data from the theoretical stellar evolution models. Statistical uncertainties were estimated using Monte Carlo simulations on the star formation history solutions. The final uncertainties presented in the SFRs were calculated by adding in quadrature the statistical uncertainties with the systematic uncertainties in SFRs. 

The distance and foreground extinction values fit using the stellar evolutionary isochrones in the program are compared to values found in the literature for all three galaxies in Table~\ref{tab:sfh_fits}. The distance moduli are in close agreement with the value determined from the TRGB by \citet{Karachentsev06}. The foreground extinction for NGC 4163 and NGC 4068 are very close to the values reported by \citet{Schlegel98}. We find a higher foreground extinction value for IC 4662 which may be a result of the low resolution of the observations by \citet{Schlegel98}.

The regions of varying stellar density studied for differing photometric completeness (\S\ref{photometry}) were also analyzed from a star formation history perspective. In two of the three galaxies, NGC 4163 and NGC 4068, the star formation history results for the regions and the global field of view were well fit by the modeled CMDs. The summation of the SFRs from the two regions was nearly identical to the SFRs found for the global data. We use the results from the single, global field of view for these two galaxies throughout the paper to calculate the duration of the starbursts in these two galaxies. The spatial information obtained from the separate regions is retained and analyzed separately in \S\ref{gvl}. The star formation history of the third galaxy, IC 4662, was more difficult to fit due to the more complex environment and higher extinction in this galaxy. We found that for IC 4662, splitting the galaxy into two regions of similar stellar density characteristics allowed the CMD fitting program to match the modeled CMD to the observed data more robustly. The higher surface brightness region was best fit by applying 0.4 mag of internal differential extinction while the lower surface brightness region matched best with a slightly lower value of 0.2 mag. The final SFRs were calculated by summing the rates from the two regions and adding their uncertainties in quadrature.

\begin{figure}
\plotone{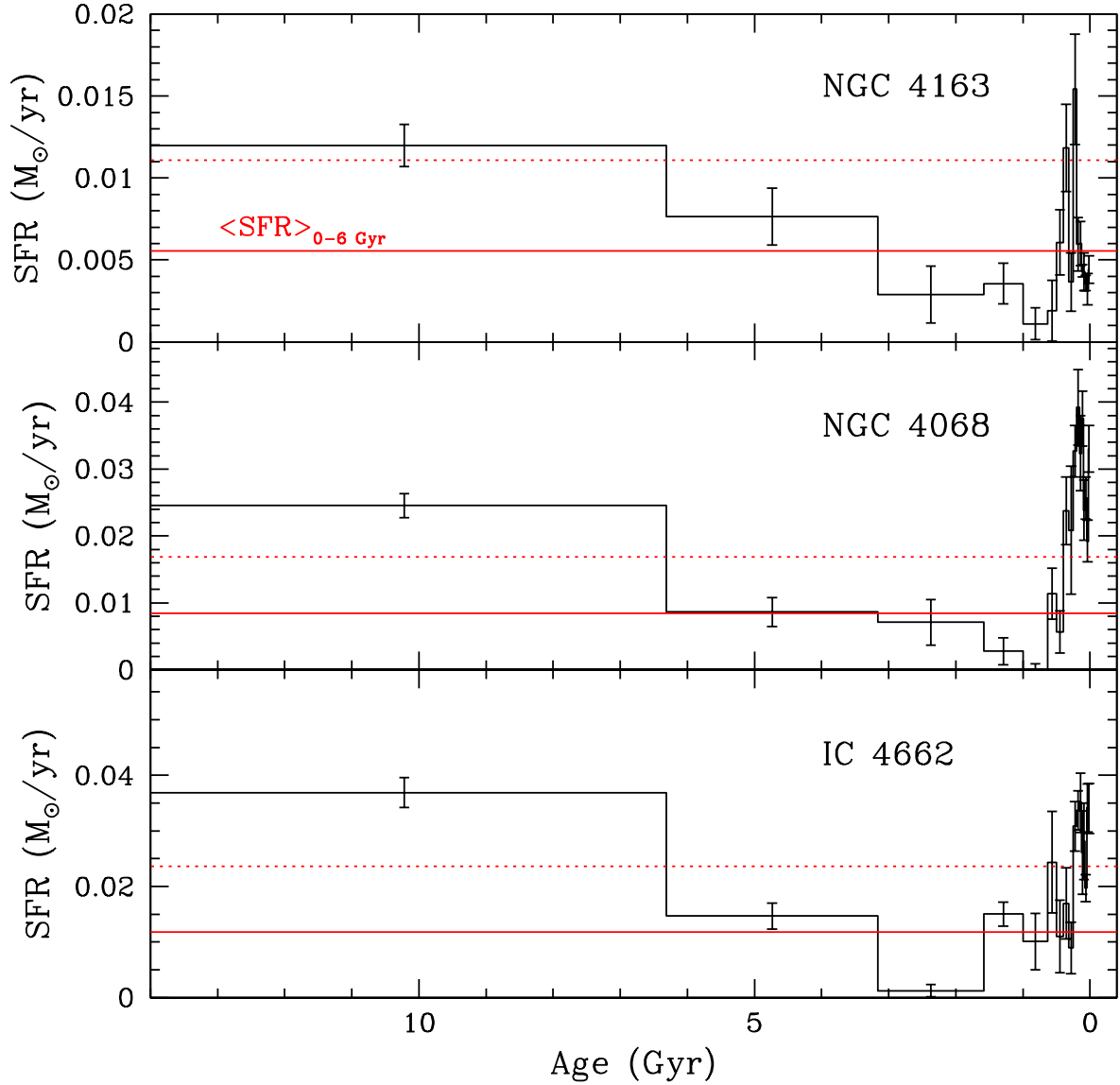}
\caption{Complete SFRs for NGC 4163, NGC 4068, and IC 4662 for the last 14 Gyr. The solid line is the average SFR for the galaxy during the last $6$ Gyr as discussed in \S\ref{duration}. The dotted line is twice this average. The final solution for IC 4662 used in the determination of the starburst duration is the summation of the SFRs from the regions of high and low stellar densities.}
\label{fig:sfrates_14Gyr}
\end{figure}

\begin{figure}
\plotone{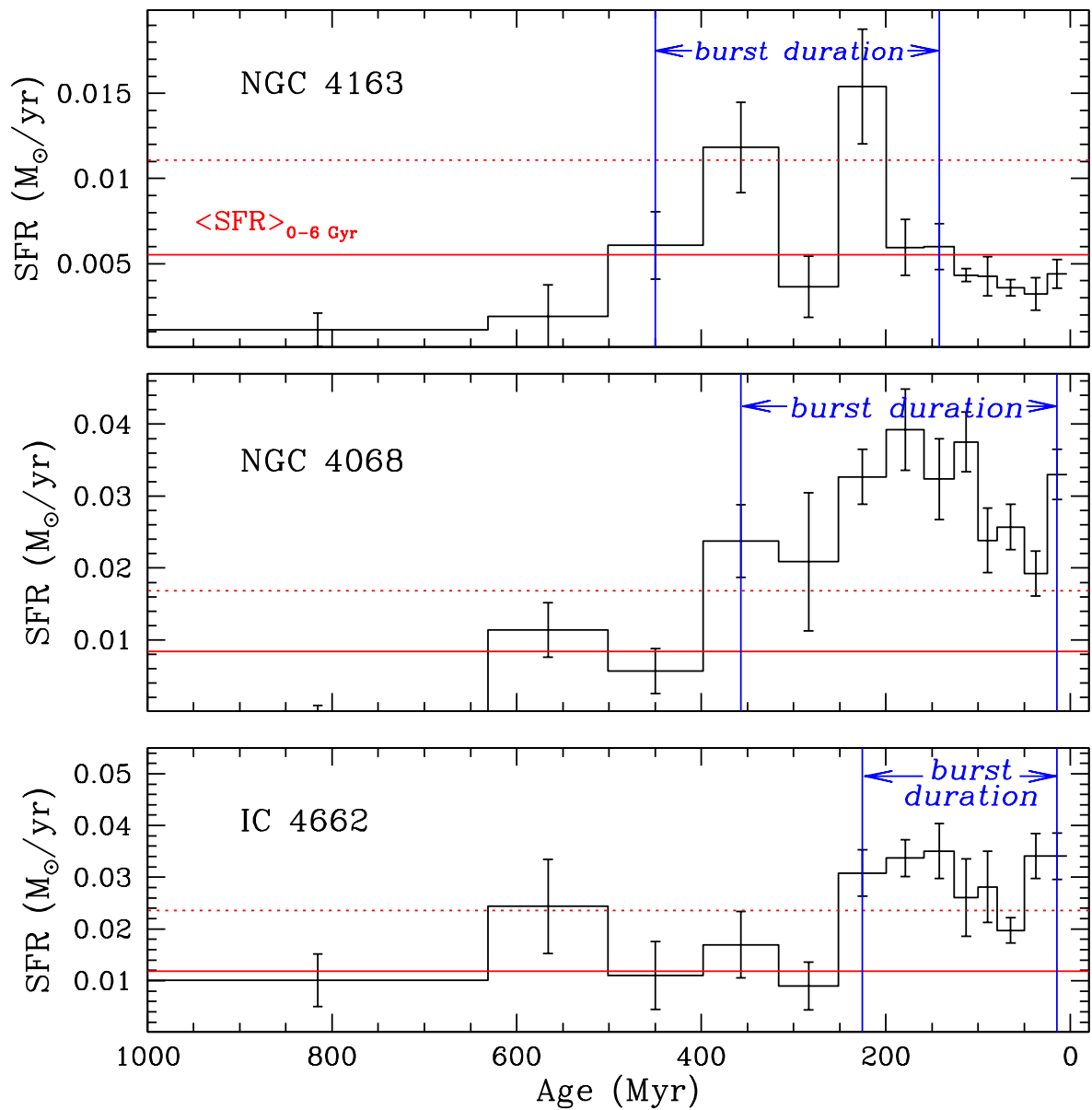}
\caption{SFRs for NGC 4163, NGC 4068, and IC 4662 for the last 1 Gyr. The solid red line is the average SFR for the galaxy during the last $6$ Gyr as discussed in \S\ref{duration}. The dotted red line is twice this average. The beginning and end of the starbursts are marked in blue vertical lines. The final solution for IC 4662 used in the determination of the starburst duration is the summation of the SFRs from the regions of high and low stellar densities.}
\label{fig:sfrates_1Gyr}
\end{figure}

The bursts for NGC 4163, NGC 4068, and IC 4662 identified in the CMDs are confirmed in the reconstruction of each galaxy's star formation history. We present the star formation histories for the lifetime of each galaxy in Figure~\ref{fig:sfrates_14Gyr} including systematic and statistical uncertainties. The ancient SFRs have been averaged over larger time bins as there is little distinction in the color-magnitude diagram between stars formed with any greater time resolution without deeper observations. The uncertainties in the older time bins are smaller due to this averaging over a longer time period. The recent SFRs are shown with a finer time resolution as these earlier times can be more precisely derived from the data. Figure~\ref{fig:sfrates_1Gyr} highlights the star formation from the past 1 Gyr only. The higher SFR in the recent time bins reflects the clustering of BHeBs identified in the CMD and delineates these dwarf galaxies as starburst systems. The durations are clearly resolved by this technique.

$\textbf{NGC 4163}$. This galaxy is the second closest blue compact dwarf (BCD) galaxy and is part of the Cane Venatici I (CVn I) cloud \citep{Karachentsev06}. It is the most spherical in morphology of the three dwarf galaxies in our sample. The star formation history in Figure~\ref{fig:sfrates_1Gyr} shows an elevated SFR in recent time bins but the burst is not ongoing in the present day in this post-starburst galaxy. In comparison with the other two galaxies, NGC 4163 has the lowest amplitude burst of the sample.

$\textbf{NGC 4068}$. This galaxy is also a BCD and member of the CVn I cloud, although it is significantly farther than NGC 4163 \citep{Karachentsev06}. There are a number of prominent star forming regions in the image and the SFR shows a ramp in  star formation activity from a very quiescent period to a burst at the present day. 

$\textbf{IC 4662}$. This isolated galaxy is the closest blue compact dwarf (BCD) galaxy and has large extended regions of star formation visible in the optical images \citep{Karachentsev06}. IC 4662 is considered a bursting galaxy based on its H$\alpha$ EW measurements \citep{Lee06} and by the SFRs derived from their H$\alpha$ spectra \citep[e.g.,][]{Hunter04, Kaisin07}. The SFRs in Figure~\ref{fig:sfrates_1Gyr} are a composite of star formation activity summed from the regions of differing stellar density in the galaxy. 

\subsection {Measuring the Durations of Three Starbursts \label{duration}}

Measuring a burst duration requires defining the beginning and ending points 
of a burst event. The profile of the SFR across time can be a guiding indicator in determining the characteristic level of a burst and dictate the bursts' temporal limits. However, because of the stochastic nature of star formation, 
precisely defining these temporal limits is somewhat problematic and 
potentially very subjective. Quantitatively, a useful parameter to consider is the birthrate parameter, \textit{b}, which compares the SFR of a given time bin to the average SFR over the life of a galaxy \citep[e.g., \textit{b} $\equiv$ SFR / $<$SFR$>_{past}$][]{Scalo86, Kennicutt98}. However, most dwarf galaxies show evidence of an elevated period of star formation early in the history of the universe during initial galaxy assembly. This has been known for some time for the gas poor dwarf spheroidals \citep[e.g.,][]{Mateo98, Grebel01}, but recent observations of the gas rich dwarf irregulars also show similar evidence \citep[e.g.,][]{Skillman03, Dolphin05}. Thus, for our purposes, a lifetime average star formation is inflated by this early epoch of star formation and is significantly higher than the average star formation rate in the current epoch. Since it is increased star formation in the current epoch that we are concerned with, we introduce a birthrate parameter for recent time epochs \textit{b}$_{recent} \equiv$ SFR / $<$SFR$>_{6-0 Gyr}$ (which excludes our earliest time bin) to study the profile of SFR and restrict our analysis to times after the initial assembly of the galaxy. The SFRs over the last 6 Gyr should be more indicative of the behavior of an unperturbed star forming disk and applies the more recent SFRs as a metric for determining the beginning and ending points of a burst.

An empirical value of $\textit{b} = 2$ has been suggested as a threshold for dwarf galaxy starbursts by \citet{Kennicutt05} while \citet{Brinchmann04} report a range of $\textit{b} = 0.1 - 30$ for a wider demographic of starburst galaxies. The value of \textit{b} has a strong redshift dependence (typical values of \textit{b} for starbursts being lower at high redshift due to the different levels of star formation activity present at those times) which cautions against choosing an absolute value of \textit{b} to denote starbursts across cosmological scales. Our galaxy sample is restricted to nearby galaxies so we will consider \textit{b} to be a definite quantity in this limited context. We compare the average SFR over the last $6$ Gyr with the SFR in recent time bins (i.e., \textit{b}$_{recent}$). Following \citet{Kennicutt05}, we use \textit{b}$_{recent} \geq 2$ to identify bursts and a threshold of \textit{b}$_{recent} \simeq 1$ to demarcate the beginning and end of the bursts. The cut-off value of $\simeq 1$ includes the ramp up time to a peak burst SFR and the ramp down to a lower, more sustainable SFR. The SFRs for \textit{b}$_{recent} = 1$ and 2 are plotted in Figure~\ref{fig:sfrates_1Gyr} with the beginning and end points of the burst marked for each galaxy. The durations, presented in Table~\ref{tab:durations}, range from $225\pm25$ to $385\pm50$ Myr. For NGC 4068 and IC 4662, the durations are lower limits since the starbursts are ongoing at the present day. NGC 4163 may be better described as a post-starburst system with the recent burst ending $\sim65$ Myr ago. This system shows a more complete SFR profile of a starburst with an elevated trailing SFR after the peak of the burst. 

The uncertainties are driven by the resolution of the time bins shown in Figure~\ref{fig:sfrates_1Gyr}. The maximum value of \textit{b} for a galaxy listed in Table~\ref{tab:durations} gives a perspective on the relative intensities of the bursts using each galaxy's unique star formation history. The durations measured were not sensitive to the different input parameters to CMD fitting program. The bursts are clearly identifiable from the BHeB and MS branches of the CMDs and are robustly derived for these galaxies. The implications of the lengths of these burst durations will be discussed in \S\ref{implications}.

\subsection {Global Bursts vs. Localized Star Clusters \label{gvl}}

\begin{figure}
\plotone{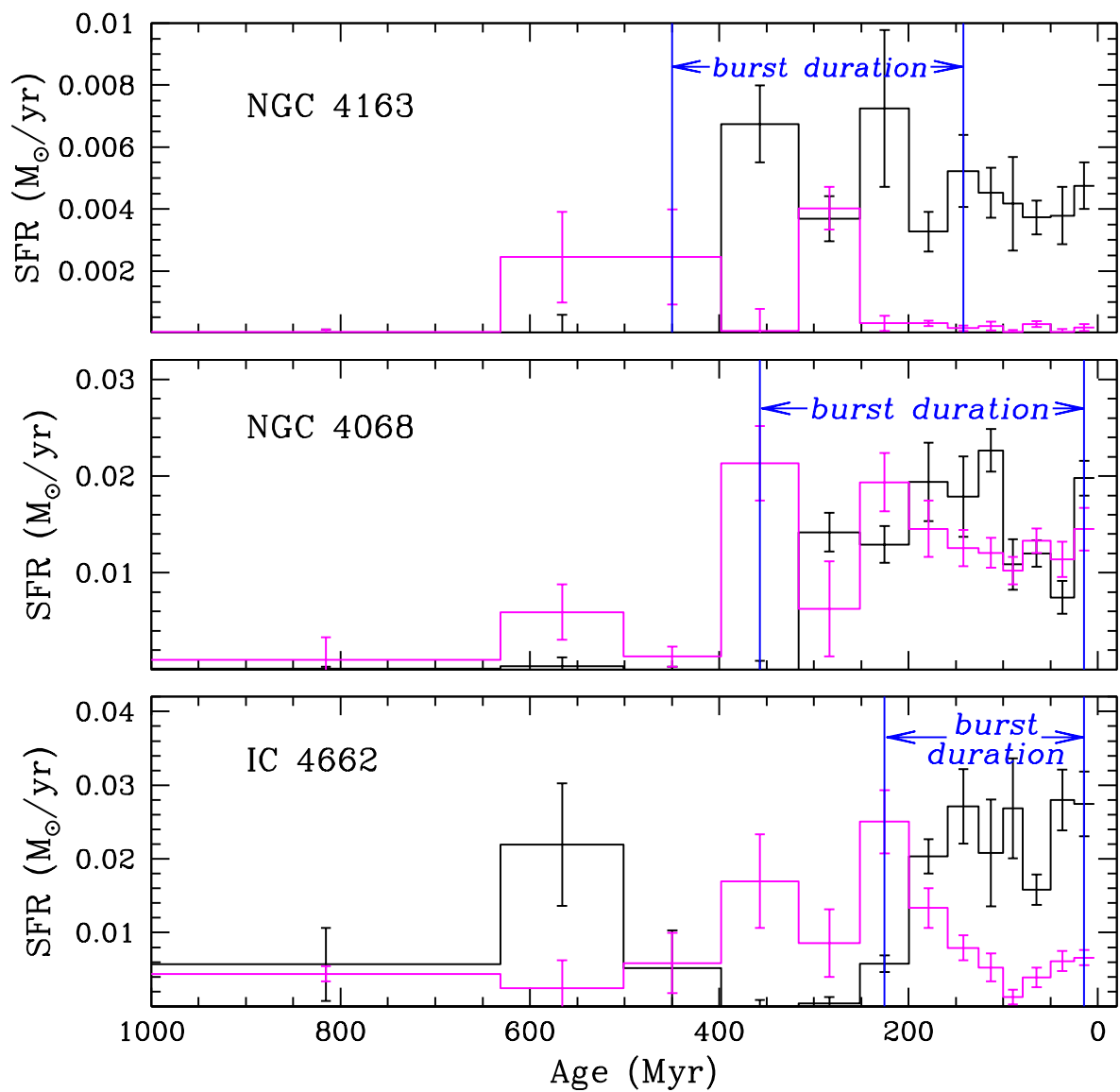}
\caption{SFRs for NGC 4163, NGC 4068, and IC 4662 from the regions of high (black line) and low (magenta line) stellar densities. The beginning and end of the starbursts are marked in blue vertical lines. Note the temporal shift of the peak SFR in the different regions. In NGC 4163 and IC 4662, the areas that appear to be bursting in the optical image show the peak SFR in more recent time bins than the non-bursting regions in the images suggesting that the burst has migrated within the galaxy over the past few hundred million years. NGC 4068 shows star formation with bursting characteristics in both regions of the galaxy in all recent time bins.}
\label{fig:sfrates_regions}
\end{figure}

The starbursts in the dwarf galaxies NGC 4163, NGC 4068, and IC 4662 are not limited to the areas that appear to be bursting at the present day in the optical images (i.e., regions of highest surface brightness and stellar density). Rather, when averaged over the last $200-400$ Myr, the bursts are more global in nature. We studied the regions where bursts \textit{appeared} to be localized in the images (i.e., the regions of higher surface brightness where we applied separate artificial star tests detailed in \S\ref{photometry}) and compared these regions' SFR to the rates in the  seemingly non-bursting areas (i.e., the regions of lower surface brightness also discussed in \S\ref{photometry}) of each galaxy. The comparisons showed that although the low surface brightness regions do not exhibit starburst characteristics in the image, they have indeed been sites of significant star formation activity in the recent past (t$<400$ Myr) as shown in Figure~\ref{fig:sfrates_regions}. In two of the three galaxies, NGC 4163 and IC 4662, the elevated star formation in the two regions lag one another, with the most recent high SFRs occurring in the most crowded, highest surface brightness regions, and with the lower surface brightness areas showing higher SFRs in the adjacent earlier time bins. In NGC 4163, the most recent SFR is concentrated entirely in the high surface brightness region of the optical images. However, there has been star formation of burst proportions in the outer regions of the galaxy $\sim300$ Myr ago, just prior to the present day activity (top panel; Figure~\ref{fig:sfrates_regions}). A similar pattern is seen in IC 4662 (bottom panel; Figure~\ref{fig:sfrates_regions}) where the lower stellar density regions show bursting levels of star formation $\sim200$ Myr ago. The third galaxy in our sample, NGC 4068, shows bursting levels of star formation in both the concentrated regions of stars and the outer regions of the galaxy in the most recent times bins without any time lag. Elevated levels of star formation do not appear to be limited to areas of high stellar density but can be pronounced in areas of low stellar density as well.

When approached from a 1 Gyr perspective, it appears the bursts have been more global in nature, migrating through different parts of a galaxy. Although the SFR may be unsustainable for the durations of $\sim 200 - \sim 400$ Myr in any particular region of a galaxy due to fuel limitations and possible feedback mechanisms, it appears that the starburst event moves around the galaxy. The burst therefore affects much more of the structure and dynamics of the host galaxy than a local phenomenon could, and is sustainable for longer duration periods of $\sim10^{8}$ yr. The idea that starbursts can be more global in nature agrees with what \citet{Cannon03} found in NGC 625. These authors report that the star formation is widespread over the past 100 Myr. Other authors report a related result that bursting star formation in starburst galaxies is not confined to star clusters but is found throughout the diffuse star formation in NGC 3310 \citep{Meurer00} and in field stars outside the central super star cluster (SSC) in NGC 1705 \citep{Annibali03}. 

One of the frequently calculated parameters for starbursts is the intensity of star formation per unit area or $\Sigma$ (\msun yr$^{-1}$ kpc$^{-1}$). It is a useful parameter because it is closely related to the gas surface mass density and because it more closely reflects local disk physics than global quantities like the total SFR \citep{Heckman05}. However, the intensity also relies on having a well-defined area for a burst. Given that the complete burst event appears to migrate around the galaxy, the intensity depends on the definition of the area, which changes during the burst. Additionally, elevated rates of star formation in the lower stellar density regions contribute to the SFR in even the most recent time bins, making a definition of area even more ambiguous. Thus, while the intensity parameter is useful in quantifying the intensity of $instantaneous$ star formation associated perhaps with super star clusters highlighted in the UV and other wavelengths, it could also be misleading for longer duration, spatially distributed events.

\section{Implications of Longer and More Global Bursts \label{implications}}

\subsection {Flickering Star Formation within a Burst \label{flickering}}

Previous measurements of $10^{6} - 10^{7}$ yr starburst durations (see references in \S\ref{intro}) are probably measuring only part of a longer starburst event. Within a starburst galaxy, there are localized regions of intense star formation that are readily studied at many wavelengths due to the intensity of emitted radiation. The duration of star formation measured in such areas can depend on the wavelength observed. For example, timescales derived using H$\alpha$ or ultraviolet emission measurements will be on the order of the lifetime of the massive stars producing the H$\alpha$ or UV radiation (i.e., $\sim10$ Myr). We briefly summarize some of the durations from the literature and the wavelength from which the duration was derived in Table~\ref{tab:prev_methods}. Many studies adopt an instantaneous burst model, assuming that the formation of thousands of massive stars within a short period of time ( $\simeq10^{6}$ yr), would disrupt any additional significant star formation by the mechanical energy injected into the area \citep[e.g.,][and references therein]{Mas-Hesse99, Mas-Hesse00}. 

However, as we have shown, the starburst phenomenon can be a longer and more global event than related by the lifetime of individual stars or pockets of intense star formation. The short duration timescales are instead a measure of the ``flickering" created by currently active pockets of star formation that move around the galaxy. Measuring the characteristics of just one of these flickers reveals much about an individual star formation region but does not measure the totality of the starburst phenomenon in the galaxy. If starbursts are indeed a global phenomenon, then the events are longer than the lifecycle of any currently observable massive star or area of intense star formation and the bursts are $not$ instantaneous. An observation that measures currently observable star formation activity will therefore measure the ``flickering" associated with a starburst pocket and not the entire phenomenon.

\subsection{Timescales and Galactic-size Bursts \label{scales}}

A basic physical property of a galaxy is the characteristic time needed to communicate an event or disturbance across the entire system. Such a timescale could be a typical rotation period (t$\sim2\pi$R$_{\textrm{e}}$/V) for a disk with solid-body rotation or a crossing time for a spheroidal system such as a star cluster (t$\sim$2R$_{\textrm{e}}$/$\sigma$). We calculate characteristic timescales of $100\pm10$ to $265\pm35$ Myr for the three galaxies under study, using radii and inclination corrected maximum rotation curve velocities from the HyperLeda database \citep{Paturel03}, as shown in Table~\ref{tab:timescales}. The burst durations we report are all greater than the characteristic timescale for the host galaxies. 

It is interesting to compare the shorter durations of the ``flickering" star formation we associate with localized star formation to the characteristic timescale for a star cluster. We estimate the timescale to be $\sim1$ Myr assuming that the effective radius of a cluster is a few pc with a velocity dispersion of $\sim$ 10 km s$^{-1}$ \citep[e.g.,][]{ Smith00, Larsen08}. This value agrees with the estimation of a few Myr reported by \citet{Heckman05} using a dynamical timescale calculation t$_{dyn}$ $\sim (G\rho)^{-1/2}$. Given the range of results shown in Table~\ref{tab:prev_methods}, the global bursts last on the order of $\sim1-10$ times the characteristic timescale for a single star cluster. Interestingly, \citet{Elmegreen00} showed that star formation in small regions (i.e., stellar clusters) occurs within one or two local crossing times while large-scale star formation in a galaxy can proceed at a slower rate with a timescale dictated by the properties of the larger system. Our findings of shorter durations and shorter timescales at the cluster level with longer global starburst durations and timescales at the galactic level support this scenario.

The longer duration, global bursts may be the dominate event shaping dwarf galaxy internal dynamics, composition, and morphology in isolated environments and certainly may play a significant role in their evolution in galaxy-rich environments where tidal interactions have been thought to dictate the evolutionary history of dwarf galaxies. This longer burst duration coupled with the observation that bursts are not localized events but spread across a significant area of a galaxy suggest that star formation in starburst galaxies may be self-regulating. It appears that the non-equilibrium energy output and mass transfer from an individual pocket of star formation may impact the local star cluster but do not shut-down the burst and the phenomenon to first order is not ``self-quenching". A long and global starburst event also has interesting implications from a chemical composition perspective. Starbursts of this magnitude may play a significant role in driving galactic superwinds. Superwinds are thought to be one mechanism responsible for expelling chemically enriched material into the IGM thereby enriching the IGM while simultaneously lowering the metallicity of the host galaxy \citep[e.g.,][]{Strickland00, Strickland04, Martin07}. 

\section{Conclusions \label{conclusions}}

We find the starburst events in NGC 4163, NGC 4068, and IC 4662 are global in nature with durations from $225\pm25$ to $385\pm50$ Myr; longer than the characteristic timescales found for each galaxy. The longer durations suggest that starbursts may not extinguish themselves through energy and mass transfer but may in fact be self-regulating. If these longer duration and more global starburst events are typical of bursting dwarf galaxies, then the starburst phenomenon may have a larger impact on galactic evolution, enrichment of the IGM with heavy metals, and chemical composition of the host galaxies than previously thought. A complete starburst event most likely includes multiple generations of star formation which requires quantifying the star formation history of the past several hundred million years to find the starburst duration. Previously reported duration timescales of $\sim10$ Myr measure the ``flickering" associated with individual star clusters within a starburst galaxy but do not measure the duration of the heterogeneous star formation present in global starburst events such as found in these three galaxies. Future work (McQuinn et al. in prep.) will extend this analysis to a larger sample of nearby bursting dwarf galaxies.

\section{Acknowledgments}
Support for this work was provided by NASA through grants AR-10945 and AR-11281 from the Space Telescope Science Institute, which is operated by Aura, Inc., under NASA contract NAS5-26555. E.~D.~S. is grateful for partial support from the University of Minnesota. J.~J.~ D.  was partially 
supported as a Wyckoff fellow. K.~B.~W.~M. gratefully acknowledges Matthew, Cole, and Carling for their support. This research made use of NASA's Astrophysical Data System and the NASA/IPAC Extragalactic Database (NED) which is operated by the Jet Propulsion Laboratory, California Institute of Technology, under contract with the National Aeronautics and Space Administration.

We would like to thank the anonymous referee for a prompt and very helpful report.

{\it Facilities:} \facility{Hubble Space Telescope}

\clearpage
\begin{deluxetable}{lllllllll}
\tablewidth{0pt}
\tablecaption{Observation Summary and Global Parameters \label{tab:galaxies}}
\tablecolumns{9}
\tablehead{
\colhead{Galaxy} 			&
\colhead{RA\tablenotemark{a}} 	&
\colhead{Decl.}				&
\colhead{HST ID}			& 
\colhead{$\lambda_{606}$}		&
\colhead{$\lambda_{814}$} 		&
\colhead{Distance\tablenotemark{b}}	&
\colhead{A$_{R}$\tablenotemark{c}} 	&
\colhead{M$_{B}$\tablenotemark{d}}	\\
\colhead{}				&
\colhead{}				&
\colhead{}				&
\colhead{}				&
\colhead{(sec)}				&
\colhead{(sec)}				&
\colhead{(Mpc)}				&
\colhead{(mag)}				&
\colhead{(mag)}
}
\startdata
NGC 4163 & 12:12:09.1s & $+$36:10:09s & GO-9771 & 1200 & 900 & $3.0\pm0.2$ & 0.052 & $-13.94\pm0.19$\\
NGC 4068 & 12:04:00.8s & $+$52:35:18s & GO-9771 & 1200 & 900 & $4.3\pm0.2$ & 0.058 & $-15.17\pm0.20$\\
IC 4662  & 17:47:08.8s & $-$64:38:30s & GO-9771 & 1200 & 900 & $2.4\pm0.2$ & 0.188 & $-15.20\pm0.17$ \\
\enddata
\tablenotetext{a}{R.A. and Decl. in J2000 coordinates}
\tablenotetext{b}{\citet[][Uncertainties in the distance are estimated to be $8\%$]{Karachentsev06}}
\tablenotetext{c}{\citet[][$\lambda_{R} = 650$nm]{Schlegel98}}
\tablenotetext{d}{\citet{deVaucouleurs91}}
\end{deluxetable}

\clearpage
\begin{deluxetable}{lcccccc}
\tablewidth{0pt}
\tablecaption{Comparison of Distance, Extinction, and Metallicity Values \label{tab:sfh_fits}}
\tablecolumns{7}
\tablehead{
\colhead{Galaxy} 			&
\colhead{(m $-$ M)\tablenotemark{a}}	&
\colhead{(m $-$ M)\tablenotemark{b}} 	&
\colhead{Total A$_{V}$\tablenotemark{c}}	&
\colhead{Foreground A$_{R}$\tablenotemark{d}}	&
\colhead{[M/H]\tablenotemark{e}}		&
\colhead{[Fe/H]\tablenotemark{f}}	\\
\colhead{}				&
\colhead{}				&
\colhead{}				&
\colhead{(mag)}				&
\colhead{(mag)}				&
\colhead{(dex,mag)}			&
\colhead{(dex,mag)}			
}
\startdata
NGC 4163 & 27.40$\pm.04$ & 27.36 & $0.03\pm0.04$ & 0.05 & -0.9,0.2 & -1.65,0.14 \\
NGC 4068 & 28.17$\pm0.04$ & 28.17 & $0.04\pm0.04$ & 0.06 & -1.0,0.1 & na \\
IC 4662: HSB\tablenotemark{g} & 26.80$\pm0.04$ & 26.94 & 0.3$\pm0.04$ & 0.19 & -1.2,0.1 &  -1.34,0.23 \\
IC 4662: LSB\tablenotemark{h} & 26.80$\pm0.04$ & 26.94 & 0.4$\pm0.04$ & 0.19 & -1.4,0.3 &  -1.34,0.23 \\
\enddata

\tablenotetext{a}{Distance Modulus best fit by the CMD fitting program. The uncertainties are lower bounds as they include only statistical uncertainties.}
\tablenotetext{b}{Distance Modulus reported by \citet{Karachentsev06}.}
\tablenotetext{c}{Foreground and internal extinction  best fit by the CMD fitting program. The uncertainties are lower bounds as they include only statistical uncertainties.}
\tablenotetext{d}{Galactic extinction at $\lambda_{R} = 650$ nm reported by \citet{Schlegel98}.}
\tablenotetext{e}{Metallicity estimated by the CMD fitting program for elements heavier than Hydrogen.}
\tablenotetext{f}{Metallicity reported by \citet{Sharina08}. The second value (mag) is the uncertainty of $V-I$ of the RGB at M$_{I}=-3.5$.}
\tablenotetext{g}{`HSB' refers to the high surface brightness region.}
\tablenotetext{h}{`LSB' refers to the low surface brightness region.}
\end{deluxetable}

\clearpage
\begin{deluxetable}{lcccc}
\tablewidth{0pt}
\tablecaption{Duration of Starbursts \label{tab:durations}}
\tablecolumns{5}
\tablehead{
\colhead{Galaxy}			&
\colhead{Peak SFR$_{\textrm{1-0 Gyr}}$}		&
\colhead{Peak \textit{b}$_{recent}$}		&
\colhead{$<SFR>_{\textrm{6-0 Gyr}}$} 		&
\colhead{Duration}			\\  
\colhead{}				&
\colhead{(\msun yr$^{-1}$)}		&
\colhead{($0-1$ Gyr)}			&
\colhead{(\msun yr$^{-1}$)}		&
\colhead{for \textit{b}=1 (Myr) } 	
}
\startdata
NGC 4163 & $0.015\pm0.003$  & $2.7\pm0.5$  & $0.0055\pm0.0002$ & $308\pm50$ \\
NGC 4068 & $0.039\pm0.006$  & $4.6\pm0.8$  & $0.0084\pm0.0006$ & $360\pm40$ \\
IC 4662  &  $0.042\pm0.004$ & $3.5\pm0.4$  & $0.012\pm0.001$   & $225\pm25$\tablenotemark{a} \\
\enddata
\tablenotetext{a}{This is most likely a double burst separated by a lower level of SF activity whose rate still exceeds the average.}
\end{deluxetable}

\clearpage
\begin{deluxetable}{llcl}
\tablewidth{0pt}
\tablecaption{Previous Methods Used to Determine Durations \label{tab:prev_methods}}
\tablecolumns{4}
\tablehead{
\colhead{Wavelength Regime} 	&
\colhead{Object(s)} 		&
\colhead{Duration}		&
\colhead{Authors}		\\
\colhead{}			&
\colhead{}			&
\colhead{(Myr)}			&
\colhead{}		
}
\startdata
\sidehead{Short duration starbursts}
Far-UV Spectra & 7 Blue Compact galaxies& $ <$10 & \citet{Fanelli88}\\
W-R Stars & 37 Wolf-Rayet galaxies & $\sim$ a few & \citet{Conti91}\\
UV Spectra \& Images, FIR & 9 Starburst galaxies & $<$10 & \citet{Meurer95}\\
Optical Spectra \& W-R Stars & 17 Starburst objects & $<7$ & \citet{Mas-Hesse99}\\
Optical Images, CMD & 1 Starburst Galaxy & $10-15$ & \citet{Annibali03} \\
Infrared Spectra & 27 Starburst galaxies & $\sim10$ & \citet{Thornley00} \\
UV Spectra & NGC 5253 & $1-8$ & \citet{Tremonti01}\\
Far-Infrared \& Radio & 2 Wolf-Rayet galaxies & 4 \& $>30$ & \citet{Yin03}\\
\sidehead{Long duration starbursts}
Optical Images, CMD & 1 Starburst galaxy &$50-100$ & \citet{Cannon03}\\
Optical Images & 1 Starburst galaxy & $\simeq100$ & \citet{Meurer00} \\
H$\alpha$ Spectra & $\sim400$ SF galaxies & $50-100$ & \citet{Lee06} \\

\enddata

\end{deluxetable}

\clearpage
\begin{deluxetable}{llcccc}
\tablewidth{0pt}
\tablecaption{Timescales  \label{tab:timescales}}
\tablecolumns{6}
\tablehead{
\colhead{Galaxy}			&
\colhead{R$_{e}$\tablenotemark{a}}	&
\colhead{V$_{\textrm{rot}}$\tablenotemark{b}}	&
\colhead{Time}	&
\colhead{Duration}			&
\colhead{Duration/Time}			\\
\colhead{}				&
\colhead{(pc)}				&
\colhead{(km s$^{-1}$)}			&
\colhead{(Myr)}				&
\colhead{(Myr)}				&
\colhead{}				
}
\startdata
NGC 4163 & $ 710 \pm 95$ & $18.3\pm0.8$ & $240\pm35$  & $308\pm50$ & $1.3\pm0.3$ \\
NGC 4068 & $1165\pm140$ & $27.0\pm0.9$ & $265\pm35$  & $360\pm40$ & $1.4\pm0.2$ \\
IC 4662  & $ 730 \pm 75$ & $46.5\pm3.1$ & $100\pm10$  & $225\pm25$ & $2.3\pm0.3$ \\
\enddata

\tablenotetext{a}{Major axis measurements and maximum rotational velocities were taken from the HyperLeda database \citep{Paturel03}. Effective radii were calculated by converting the major axis value to radial size using the TRGB distance by \citet{Karachentsev06}.}
\tablenotetext{b}{Maximum rotational velocity corrected for inclination from the HyperLeda database \citep{Paturel03}.}
\end{deluxetable}


\clearpage

\clearpage

\clearpage

\clearpage

\clearpage

\clearpage

\clearpage

\clearpage

\clearpage

\clearpage

\clearpage

\clearpage

\clearpage

\clearpage

\end{document}